\documentclass[12pt]{article}
\usepackage[latin1]{inputenc}
\usepackage[dvips]{epsfig,rotating}
\usepackage{epic, amssymb, amsmath,latexsym,exscale,a4}

\author{
  Klaus Blindert\footnote{klaus.blindert@web.de}
}
\title{Sex and hermaphroditism in the Penna model\\
	{\large Lecture "Computer Simulation": D.Stauffer}
}

\begin{document}

\maketitle
\section{Comparing sex and hermaphroditism}

I compared sexual(SX) and hermaphroditic(HA) reproduction schemes
within the Penna model$[1]$. A fixed set of parameters was used, to make
the results comparable with recent simulations$[2]$. I also reproduced
the results for a self-emerging dominance$[2]$ in an independant simulation
for both hermaphroditic and sexual cases. A variation on partner selection
was also implemented, this shifted stable population sizes slightly.

The stable population is considered the most significant result for these comparisons
since Stauffer et al.$[3]$ gave strong argument, that from two species competing
for the same resources the one with a higher stable population while not
competing will prevail, while the other one will die out.

The parameters were: Birthrate $B=4$, minimum reproduction age $R=8$, 
mutational threshold $T=3$, and mutation rate $M=1$.
Different values for $N_{max}$ have been simulated. The starting population
was choosen to $N(0) = N_{max}/100$.

In order to get better statistics several runs have been made with a different
seed value for the random number generator, when computer time was available.

\section{Constant dominance}

To take the effect of dominant and recessive genes into account, a dominance 32 bit long
string was generated for the whole population with $d=6$ randomly set bits.
Deletrious mutations only take effect if either both genome strings have the according
bit set or one string has it set and the bit is marked as dominant by the dominance string.

As can be seen in table \ref{simple:sim:pop} the hermaphroditic populations consistantly
reached a stable population of $ N_0/N_{max}  \approx 0.23 $ whereas
the sexually reproducing populations reached $ N_0/N_{max}  \approx 0.17 $.
No significant difference  was found for longer simulation time or larger populations.

\begin{figure}
\center{\includegraphics[width=0.7\textwidth,angle=-90]{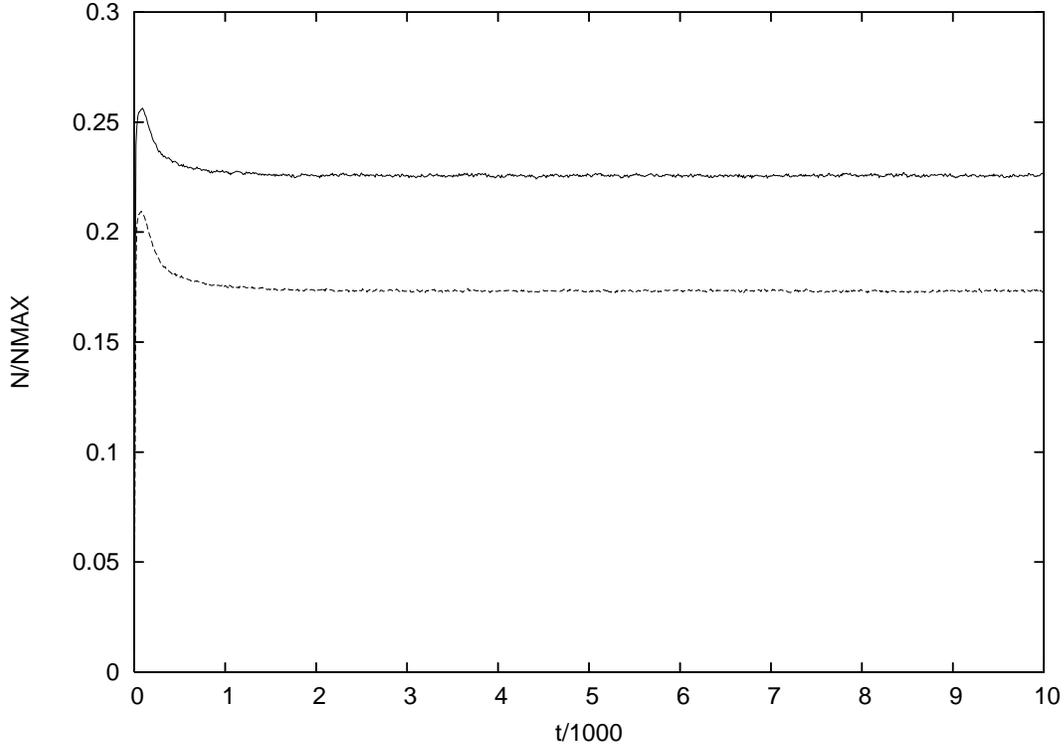}}
\label{simple:pop:compare}
\caption{
	Comparing sexual (top curve) and hermaphroditic (bottom curve) reproduction
	schemes. The dominace was assumed to be constant for the whole population
	for these simulations.
}
\end{figure}

\begin{figure}[ht]
\begin{centering}
\begin{tabular}{|c|c|c|c|c|c|}
\hline

SX/HA
& $N_{max}$ & $N_{runs}$ & Time &  $N_0/N_{max}$ \\
\hline
HA & $5 \cdot 10^5 $ & 1 & $5 \cdot 10^4 $ & $0.226$ \\
HA & $5 \cdot 10^5 $ & 10 & $1 \cdot 10^4 $ & $0.226$ \\
SX & $5 \cdot 10^5 $ & 1 & $5 \cdot 10^4 $ & $0.174$ \\
SX & $5 \cdot 10^5 $ & 10 & $1 \cdot 10^4 $ & $0.173$ \\
\hline
\end{tabular}
\label{simple:sim:pop}
\caption{
	Comparing sexual(SX) and hermaphroditic(HA) reproduction
}
\end{centering}
\end{figure}

\section{Dominance self emergence}

\begin{figure}
\center{\includegraphics[width=0.7\textwidth,angle=-90]{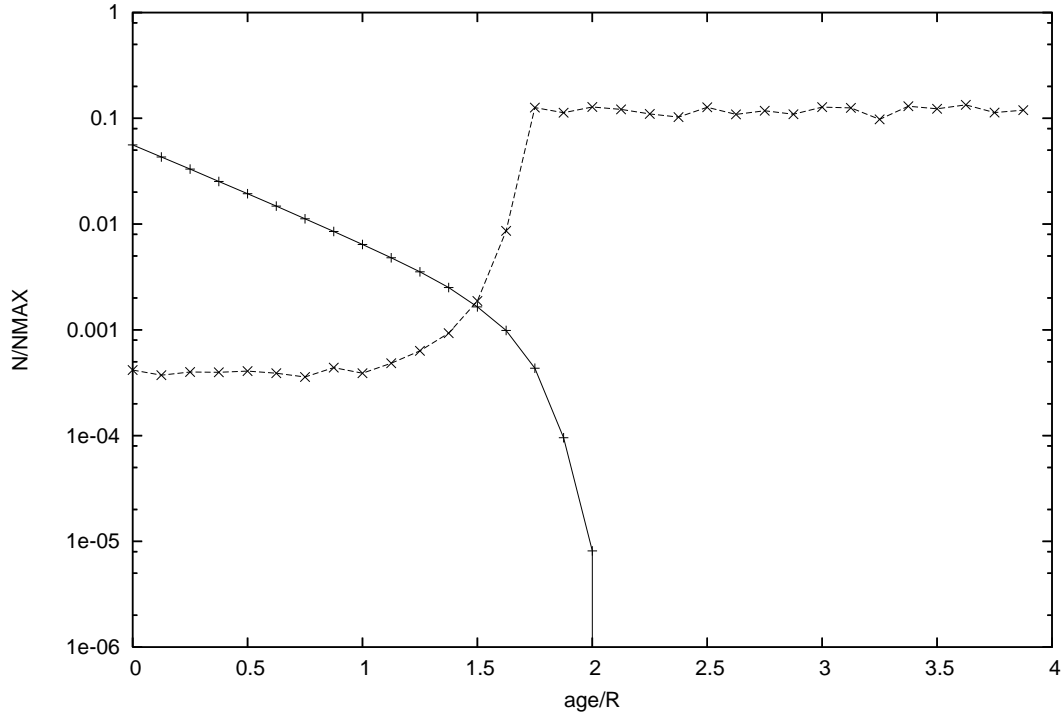}}
\label{sim4:agedom}
\caption{
	(\textbf{x}) is the number individuals with a dominance bit set at that age,
	(\textbf{+}) is the total number of individuals at that age
	(H,\texttt{J}, $N_0\approx 0.23$)
}
\end{figure}

I started from the standard Penna model with sexual and hermaphroditic
reproduction. Each individual now has a dominance bit string attached.
With every reproduction one bit in this string is flipped with a
probability of $p = 0.01$. This dominance string is then used to determine
the activity of a deletrious mutation just as in the constant dominance model.
As in recent simulations$[2]$ the number dominant bits $n_{dom}$ reached a stable value after
$t \geq 10^5 $ time steps (figure 6
) and has an age distribution
reflecting the selection pressure gradient (figure \ref{sim4:agedom}
).

\begin{figure}
\center{\includegraphics[width=0.7\textwidth,angle=-90]{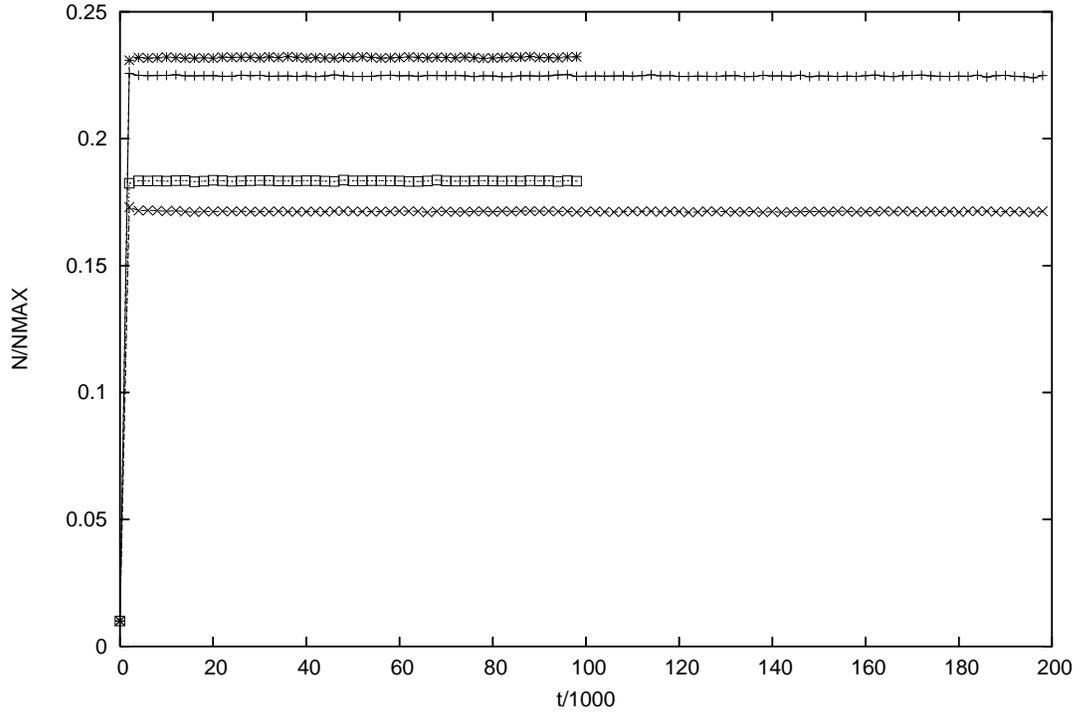}}
\label{pop:compare}
\caption{
	From top to bottom:
	a.) Hermaphroditic reproduction, J 
	c.) Hermaphroditic reproduction, 20T 
	b.) Sexual reproduction, J 
	d.) Sexual reproduction, 20T 
}
\end{figure}

\begin{figure}
\begin{centering}
\begin{tabular}{|c|c|c|c|c|c|c|}
\hline
 
SX/HA
& $N_{max}$ & $N_{runs}$ & Part.sel. & Time &  $N/N_{max}$ & $n_{dom}$\\
\hline
SX & $10^7$ & 10 & 20T & $2\cdot 10^5$ &  $ 0.17 $ & $ 8.5 $ \\
SX & $5 \cdot 10^5$ & 50 & 20T & $2\cdot 10^5$ & $ 0.17 $ & $ 8.2 $  \\

HA & $5 \cdot 10^5$ & 50 & 20T & $2\cdot 10^5$ & $ 0.23 $ & $ 8.46 $ \\
HA & $5 \cdot 10^5$ & 50 & J & $10^5$ & $ 0.23 $ & $ 9.27 $ \\

SX & $5 \cdot 10^5$ & 50 & J & $10^5$ & $ 0.18 $ & $ 8.35 $ \\
\hline
\end{tabular}
\label{sim:dom}
\caption{
	Dominance emergance simulation parameters and results
}
\end{centering}
\end{figure}


\subsection{Partner selection style}

In recent simulations$[3]$ each female or hermaphroditic individual with $age \geq R$
trys to find a fitting partner from the whole population 20 times. If no partner
is found, the individual chooses to not reproduce that time step. The simulations
with this reproduction style are marked 20T.
\clearpage
Another model was simulated, where each turn a pool of males in the apropiate
reproduction age is generated. Each female chooses one from this pool and removes
it from this pool. So the individuals form couples each turn. This model is marked as
J according to a german saying that each jacket will find fitting trousers.

However this different partner selection style only increased the stable
population sizes slightly.

\subsection{Population}

As can be seen in figure 4 
in the hermaphroditic case
a stable population of $N_0/N_{max}\approx 0.23$ was reached,
whereas the sexual reproduction scheme reached $N_0/N_{max}\approx 0.17$,
The different partner selection styles simulated did not produce important differences.

\begin{figure}
\center{\includegraphics[width=0.7\textwidth,angle=-90]{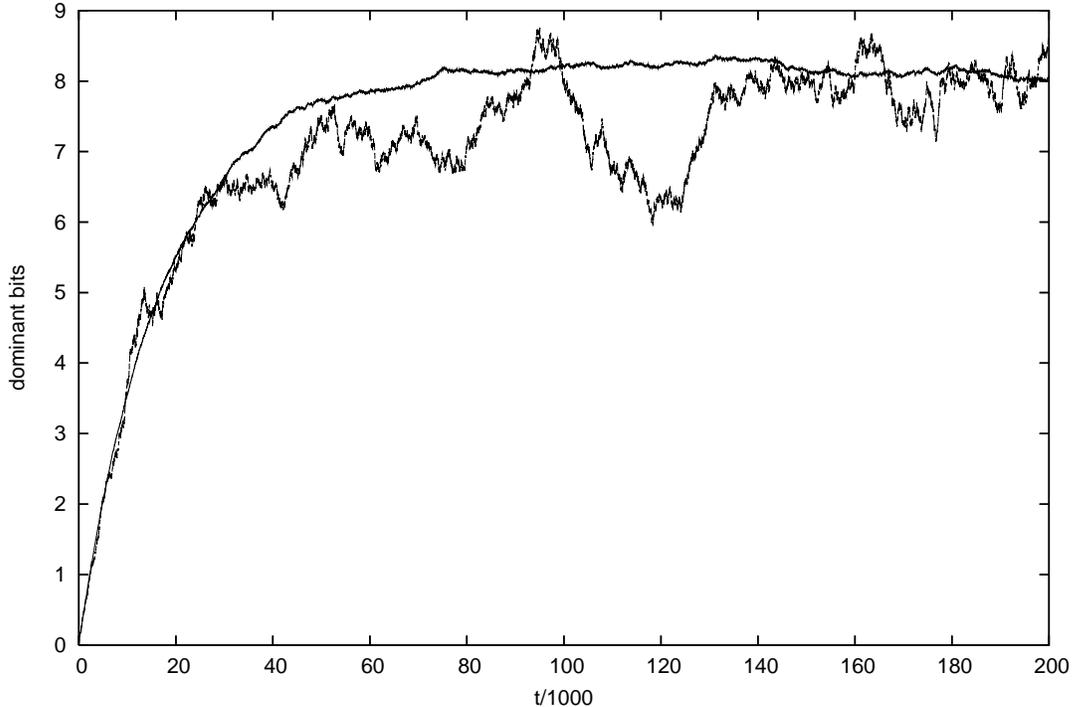}}
\label{sim2:ndombits}
\caption{
	Number of dominant bits averaged over the whole population. The fluctuating
	curve represents one run, whereas the smooth curve results from
	averaging over 50 runs. (SX,20T)
}
\end{figure}

\section{Summary}

All simulations consistantly showed that hermaphroditism reaches a higher stable
population size than sexual reproduction. The Penna model thus does not explain
the evidence found in nature that sexual reproduction offers severe benefits
over hermaphroditic reproduction. This known feature of the Penna model has
been confirmed.

Several researchers have tried to give an explanation for sexual reproduction
in comparison to asexual reproduction. Other reproduction schemes (especially
meiotic parthenogenesis) have also been analysed. See chapter 2.5.5 in 
\texttt{Evolution, Money, War and Computers} by Oliviera, Oliviera and Stauffer $[5]$.

To further compare reproduction schemes one should investigate the applicability
of these approaches to hermaphroditism.



%

\subsection{References}

\begin{tabular}{rl}
$[1]$	& \small{Penna, A bit-string model for biological aging.} \\
	& \small{J. Stat. Phys. 78, 1629 (1995)} \\
$[2]$   & \small{Garncarz, Cebrat, Stauffer and Blindert, Why are diploid genomes widespread?} \\
$[3]$ 	& \small{Stauffer, Martins and Oliviera, On the uselessness of men.}\\
	& \small{Int. J. Mod. Phys. C, 1305 (2000)} \\
$[4]$	& \small{Stauffer, de Oliviera, Moss de Oliviera, Penna and Martins}\\
	& \small{Computer simulations for biological aging and sexual reproduction}\\
	& \small{An. Acad. Bras. Ci. (2001) 73 (1)}\\
$[5]$	& \small{Moss de Oliviera, de Oliviera, Stauffer, Evolution, Money, War and Computers. }\\
	& \small{Teubner Texte, p. 60ff (1999) }\\
\end{tabular}

\end{document}